\documentclass{kapproc} 
\usepackage{epsfig}

\newcommand{\Msun}{{M_\odot}}
\newcommand{\Lsun}{{L_\odot}}

\newcommand{\mdot}{{\dot{M}}}

\kluwerbib

\begin{document}

\articletitle[\bf The chemical evolution of the solar neighbourhood]{The chemical evolution
of the \\solar neighbourhood}

\author{Dany Vanbeveren and Erwin De Donder}
\affil{Astrophysical Institute, Vrije Universiteit Brussel, Pleinlaan 2,
1050 Brussels, Belgium}
\email{dvbevere@vub.ac.be, ededonde@vub.ac.be}


\begin{abstract}
Recent models of galactic chemical evolution account for updated
evolutionary models of massive stars (with special emphasis on stellar winds) and for the
effects of intermediate mass and massive binaries. The results are summarised. We also
present a critical discussion on possible effects of stellar rotation on overall galactic chemical
evolutionary simulations.          
\end{abstract}

\section{Introduction}
\noindent Most of the chemical evolutionary models (CEMs) consider the enrichment of the
interstellar medium (ISM) by single stars only (with some parametrisation to account for the effects of
type Ia supernovae). However, observations reveal that a significant fraction of all stars has a
close companion (Mermilliod, 2001, Mason et al., 2001 and
references therein) and a few years ago we started an extended study on the effects of binaries on
CEMs. First results were published in De Donder and Vanbeveren (2002) and are summarised in
Sect.~\ref{C}.

CEMs of galaxies rely on stellar evolution and therefore improvements in the physics of the latter may
have important consequences for the former. To illustrate, Sect.~\ref{A} considers recent updates of
the  Wolf-Rayet (WR) type stellar wind mass loss in massive stars and the effect on stellar evolution
(remind that WR stars are massive hydrogen deficient core helium burning stars, further abbreviated as
CHeB). The consequences for CEM simulations will be discussed in Sect.~\ref{BH}.  

Meynet and Maeder (2000) concluded that stellar rotation modifies all the outputs of stellar evolution
of massive stars (see also Heger et al., 2000). In Sect.~\ref{G} we present a
critical discussion and we estimate the effect of rotation on CEMs.  

\section{Interacting binaries and the effect on chemical evolution}\label{C}

\noindent To study the effects of binaries on CEMs, it is essential to
combine a star formation model, a galaxy formation model, a galactic evolutionary model and a
population number synthesis (PNS) model which accounts for single stars and interacting binaries. A
description of our PNS code can be found in Vanbeveren et al. (1998a, b, c), Vanbeveren (2000, 2001),
De Donder and Vanbeveren (2002). In its present state it allows to explore a population of binaries
with initial parameters 3 $\le$ M$_{1}$/$\Msun$ $\le$ 120 (M$_{1}$ = primary mass), 0 $<$ q $\le$ 1
(q = binary mass ratio = mass secondary/mass primary) and 1 day $\le$ P $\le$ 10 years (P = binary
orbital period), and a population of single stars with 0.1 $\le$ M/$\Msun$ $\le$ 120. It accounts for
the RLOF, mass transfer and mass accretion in case A and case Br binaries, the common envelope (CE)
process and spiral-in in case Bc and case C binaries, the CE and spiral-in in binaries with a compact
companion (a white dwarf, a neutron star  or a black hole) and the effects of an asymmetric SN
explosion on the binary parameters when one of the binary components explode. We use a detailed set of
stellar evolutionary calculations for different initial metallicities (0.001 $\le$ Z $\le$ 0.02) where,
in the case of massive stars, the most recent stellar wind mass loss rate formalisms are implemented
(e.g. Sect.~\ref{A} for the WR-$\mdot$ formalism). Type Ia supernovae (SN Ia) are associated with the
single degenerate scenario (Hachisu et al., 1996)) and, separately, with the double degenerate scenario
(Iben and Tutukov, 1984; Webbink, 1984)) of mass accreting white dwarfs (notice that the only
realistic way to study the effects on CEMs of the Fe-enrichment of SN Ia's is to link a detailed PNS
model with both scenario's). All our evolutionary calculations rely on the core overshooting formalism
proposed by the Geneva group (Schaller et al., 1992).

The results of the CEM simulations with binaries performed in Brussels have been published in De
Donder and Vanbeveren (2002). In the latter paper we focussed on the elements $^{4}He$,
$^{12}C$, $^{16}O$, $^{20}Ne$, $^{24}Mg$, $^{28}Si$, $^{32}S$, $^{40}Ca$ and $^{56}Fe$. Summarising:

\begin{itemize}

\item
although binaries return less matter to the ISM than single stars, the temporal evolution of the
rate of star formation only marginally depends on whether binaries form or not

\item
the age-metallicity relation (AMR) shows the time evolution of the ratio [Fe/H] which is usually taken
as a measure of the metallicity of the galaxy. We concluded that the AMR predicted by a theoretical
galactic model hardly depends on whether binaries are included or not
 
\item
the observed solar carbon abundance and the observational behaviour of the ratio [C/Fe] as function
of [Fe/H] in the solar neighbourhood may be indirect evidence for the presence of a significant
intermediate mass binary population during the whole evolution of the solar region

\item
the temporal evolution of the elements $^{4}He$, $^{16}O$, $^{20}Ne$, $^{24}Mg$, $^{28}Si$, $^{32}S$
and $^{40}Ca$ predicted by a CEM where the effect of binaries is considered in detail differ by no
more than a factor of two to three from the results of models where all stars are treated as single
stars and where the effect of binaries is simulated only to account for the SN Ia population.   

\end{itemize}
 
The latter obviously means that as long as observational and theoretical uncertainty is at least a
factor two, the sophisticated CEMs including binaries are as good (or as bad) as those without
binaries.

\section{A WR-type $\mdot$-formalism and the effect on massive star evolution}\label{A}

\paragraph {\bf The formalism}

\noindent 
Using a hydrodynamic atmosphere code where the stellar wind is assumed to be homogeneous, 
Hamann et al. (1995) determined $\mdot$-values for a large number of WR stars. Since then evidence has
grown that these winds are clumpy and that a homogeneous model overestimates $\mdot$, typically by a
factor 2-4 (see Hamann and Koesterke, 1998, and references therein).  In the period
1997-1998 massive single star and binary evolutionary calculations were published (Vanbeveren et al.,
1998a, b, c) where we adopted the following WR mass loss rate formalism. We assumed a (simple)
relation between $\mdot$ ($\Msun$/yr) and the stellar luminosity L (in$\Lsun$)

\begin{equation} log(- \mdot) = a log L - b			\end{equation} and we tried to
find appropriate values for the constants a and b by accounting for the following criteria and
observations which were available in 1998:

\begin{itemize}
\item
the WN5 star HD 50896 (WR 6) has a luminosity log L = 5.6-5.7 and a log(-$\mdot$) $\approx$ -4.4
(Schmutz, 1997)
\item
the log(-$\mdot$) of the WNE component of the binary V444 Cyg (WR 139) derived from the observed
orbital period variation is $\approx$ -5 (Underhill et al., 1990). Its orbital
mass is 9 $\Msun$ and using a mass-luminosity relation holding for WNE-binary components (Vanbeveren and
Packet, 1979; Langer, 1989) it follows that its log L = 5 
\item
the observed masses of black hole (BH) components in X-ray
binaries indicate that stars with initial mass $\ge$ 40 $\Msun$ should end their life with a mass larger
than 10 $\Msun$ (= the mass of the star at the end of CHeB) 
\item
the WN/WC number ratio predicted by stellar evolution
depends on the WR-type $\mdot$-formalism. Therefore, last but not least, we looked for a and b values
which predict the observed WN/WC number ratio ($\approx$ 1) for the solar neighbourhood.
\end{itemize}

This exercise allowed us to propose the following relation:

\begin{equation}
log (-\mdot) = log L - 10
\label{eq:wolf}  		
\end{equation}
                         				     
Since then more galactic WR stars have been investigated with detailed atmosphere codes including the 
effects of clumping. They are listed in Van Bever and Vanbeveren (2002) where we conclude that
Equation~\ref{eq:wolf} still fits fairly well these new observations. Interestingly, the WC6 star
OB10-WR1 in the association OB10 of M31 has been investigated by Smart et al. (2001) and also fits
Equation~\ref{eq:wolf}.

Nugis and Lamers (2000, NL) proposed an alternative WR mass loss relation. It can easily be checked
that when the observed $\mdot$ is compared to the predicted ones for the WR stars mentioned above, NL is
not more accurate than our relation (see Van Bever and Vanbeveren, 2002). 

\paragraph {\bf The effect of metallicity Z on $\mdot$}

The Z-dependency of WR mass loss rates deserves some attention. Nugis and Lamers
adopt a $\mdot$-relation which is Z-dependent where, as usual, Z is defined as the abundance of all
elements heavier than helium. This means that they intrinsically assume that WN and WC stars with the
same luminosity and belonging to the same stellar environment (thus having a similar
heavy-metal-abundance) may have quite a different mass loss rate (remind that WN and WC stars have CNO
abundances which differ by about a factor 100). However, there is no observational evidence that this is
the case. 

It is generally accepted that the WR-type stellar wind is radiation driven and therefore one may expect
that {\it the heavy metals} are the main drivers. On that line of argument it is more plausible that the
$\mdot$ depends roughly on some power $\zeta$ of the heavy-metal-abundance of the WR (which is roughly
proportional to the heavy-metal-abundance of the WR progenitor population). 

Hamann and Koesterke (2000) investigated 18 WN stars in the LMC. When these data are plotted in a 
$\mdot$-L diagram, there seems to be no obvious relation. However, as an exercise, Vanbeveren (2001)
used a linear relation similar to (Equation~\ref{eq:wolf}) and shifted the line until the sum of the
distances between observed values and predicted ones was smallest (a linear regression where the slope
of the line is fixed). This resulted in a shift of -0.2 which, accounting for the
heavy-metal-abundance of the Solar neighbourhood and of the LMC, corresponds to $\zeta$ = 0.5. To
be more conclusive, many more observations are needed where detailed analysis may give reliable stellar
parameters, but from this exercise it looks as if {\it WR mass loss rates are
heavy-metal-abundance-dependent}.

\paragraph{The effects on massive star evolution}\label{B}

\begin{figure}[tb]
\begin{center}
\includegraphics[width=9cm]{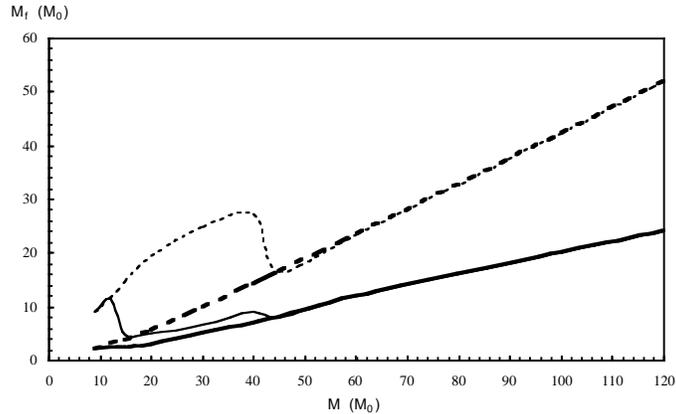}
\end{center}
\caption[]{The final mass $M_{f}$ (in $\Msun$) of the star as a function of the initial mass M (in
$\Msun$) of the star. Full lines correspond to Z=0.02 (solar neighbourhood), dashed lines
correspond to
Z=0.002 (SMC). Thick lines hold for interacting binary components, thin lines hold for single
stars.} 
\label{fig:finalmass}
\end{figure}

\noindent Due to stellar wind mass loss and/or due to Roche lobe overflow (RLOF), a massive star may 
become a hydrogen deficient CHeB star. The further evolution is governed by the WR-type
$\mdot$. The effects on massive star evolution have been studied by Vanbeveren et al.(1998 a, b, c)
(see also Vanbeveren, 2001; Van Bever and Vanbeveren, 2002). We distinguish two effects:

\begin{itemize}
     
 \item Figure~\ref{fig:finalmass} gives the final mass of the star prior to core collapse, for single
stars and for interacting binary components, for Z=0.02 (solar neighbourhood) and for Z=0.002
(appropriate for the SMC), assuming that the WR mass loss rates are
heavy-metal-abundance-dependent.  We conclude that {\it with our preferred WR mass loss
formalism, stars with initial mass between 40 $\Msun$ and 100 $\Msun$ end their life with a mass
between 10
$\Msun$ and 20 $\Msun$}, much larger than the pre-supernova masses predicted by evolutionary
calculations which were considered as a standard in the year 1997-1998 and even later. Notice that with 
Equation~\ref{eq:wolf} it becomes straightforward to explain the large masses of the BH components in
the high mass X-ray binary Cyg X-1 and in a number of low mass X-ray binaries.

\item  The $\mdot$ during CHeB significantly affects the evolution of the convective CO-core (we
further use simply the CO-core). At the onset of CHeB, the CO core rapidly increases in mass.
When the
$\mdot$ is large (as is the case for stars with an initial mass $\ge$ 40 $\Msun$ when
Equation~\ref{eq:wolf} applies), the extend of the CO-core reaches a maximum and then starts
to decrease. The layers outside the decreasing CO core are carbon rich and remain so till the
star's death. However, when the $\mdot$ is small (as is the case when the heavy-metal-abundance is
low and the
$\mdot$ depends on the heavy-metal-abundance as discussed in the precious paragraph), the mass of the
 CO core increases till the end of CHeB. At the end of the star's life, this CO
core mainly consists of O. Notice that this is also the type of CO-core evolution in massive stars which
retain their hydrogen rich envelope, i.e. in single stars where the stellar wind mass loss is too small
to remove the hydrogen rich layers. 

\end{itemize}   

In Sect.~\ref{BH} we will show that the effects discussed above may be very important for chemical
evolutionary simulations of the early phases of a galaxy when the heavy-metal-abundance was small.
     
\section{Black hole formation and the effect on chemical evolution}\label{BH}

\noindent Accounting for the WR-type $\mdot$-values (Sect.~\ref{A}),
stellar evolution in combination with the supernova explosion simulations of Woosley and Weaver (1995)
predicts that low mass BHs are descendants from single stars with an initial mass between $\sim$20-25
$\Msun$ and $\sim$35-40 $\Msun$, whereas all stars with an initial mass $\ge$ 40 $\Msun$ form high mass
BHs. De Donder and Vanbeveren (2003) investigated the effects of stars with an initial mass $\ge$ 40
$\Msun$ on chemical evolution simulations of the early Milky Way (early means the initial phase in the
evolution of the Milky Way where the chemical enrichment of the ISM was due to massive stars only,
roughly the phase where [Fe/H] $<$ -1). Here we summarise some results.

One of the crucial questions in order to understand the early evolution of a galaxy is whether or not
all BHs are preceded by a SN explosion. The last few years a class of extremely energetic supernova
has been recognised (Nomoto et al., 2001) and it is tempting to associate them with the formation
of a BH (Woosley, 1993; Paczynski, 1998; Iwamoto et al., 1998). The amount of
ejected $^{56}Ni$ that is needed to explain these hypernova is very large, as large as
0.4-0.7 $\Msun$ (Sollerman et al., 2000). This means that if the formation of all BHs would be
preceeded by a hypernova, their contribution to the Fe enrichment of the ISM could be
substantial. 

A second crucial question for the early galactic evolution, related to the association of BHs and
hypernovae, is whether or not the WR-like $\mdot$-values in BH-progenitors are heavy-element-dependent
or not. As outlined in Sect.~\ref{A}, when the $\mdot$ is large and when the BH
formation is preceded by a supernova/hypernova, the mass lost by stellar wind during CHeB and the
ejected matter prior to the BH formation will contain large amounts of carbon. When the $\mdot$ is
small, obviously the mass lost by stellar wind during CHeB is small but the ejected matter prior to BH
formation will contain large amounts of oxygen and a smaller amount of carbon.

\begin{figure}[tb]
\begin{center}
\includegraphics[width=9cm]{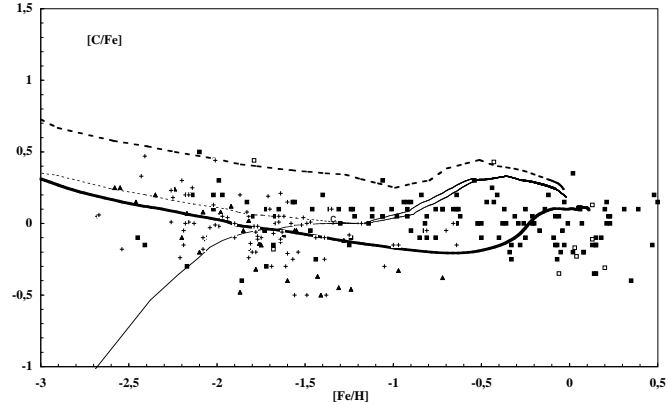}
\end{center}
\caption[]{The simulated [C/Fe]-[Fe/H] relation compared to observations.} 
\label{fig:CFe}
\end{figure}

\begin{figure}[tb]
\begin{center}
\includegraphics[width=9cm]{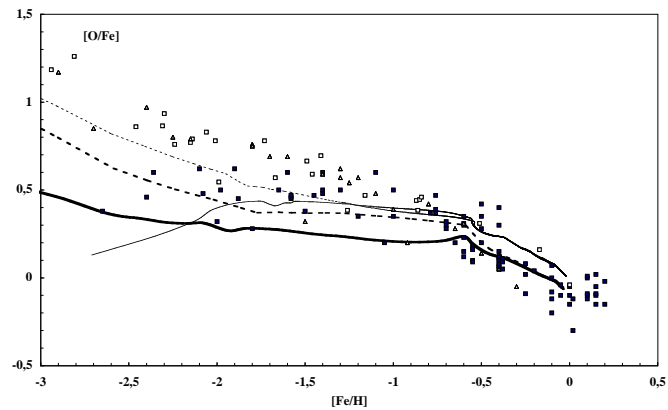}
\end{center}
\caption[]{The simulated [O/Fe]-[Fe/H] relation compared to observations.} 
\label{fig:OFe}
\end{figure}

We simulated the early galactic chemical evolution of the solar neighbourhood for the
following scenario's:

\begin{itemize}
\item all the stars with an initial mass $\ge$ 40 $\Msun$ directly collapse into a
massive BH without the ejection of matter (scen a, bold full line in Figure~\ref{fig:CFe} and
Figure~\ref{fig:OFe})

\item all the stars with an initial mass $\ge$ 40 $\Msun$ explode as a hypernova with
the ejection of 0.7 $\Msun$ of $^{56}Ni$ = 0.7 $\Msun$ of Fe (i.e. all BHs are
preceded by a hypernova) (scen b, thin full line in Figure~\ref{fig:CFe} and Figure~\ref{fig:OFe})

\item stars with an initial mass $\ge$ 40 $\Msun$ form BHs with matter ejection, the
ejected mass contains little $^{56}Ni$, the WR-type $\mdot$-formalism is not heavy-element-dependent
but satisfies Equation~\ref{eq:wolf} so that the mass lost by stellar wind during CHeB and the ejected
mass prior to the BH formation contains large amounts of carbon (scen c, dashed line in
Figure~\ref{fig:CFe} and Figure~\ref{fig:OFe})

\item stars with an initial mass $\ge$ 40 $\Msun$ form BHs with matter ejection, the
ejected mass contains little $^{56}Ni$, the WR-type $\mdot$-formalism is heavy-element-dependent
and/or the single star mass loss rates are not large enough to remove the hydrogen rich layers so that
the mass lost by stellar wind during CHeB and the ejected mass prior to the BH formation contains only a
small amount of carbon but large amounts of oxygen (scen d, dotted line in Figure~\ref{fig:CFe} and
Figure~\ref{fig:OFe}).

\end{itemize}

When we compare our simulated [O/Fe]-[Fe/H] and [C/Fe]-[Fe/H] relations to observations
(Figure~\ref{fig:CFe} and Figure~\ref{fig:OFe}), we conclude:

\begin{itemize}

\item with scen a the predicted [O/Fe] is too small 
\item with scen b, iron increases very rapidly. As a consequence the predicted [O/Fe] and [C/Fe] is
always much too small, independent from the adopted WR-type $\mdot$-formalism
\item with scen c the predicted [C/Fe] is too large; the predicted [O/Fe] is larger than with scen a but
it is still rather low compared to observations
\item scen d is by far the best model.

\end{itemize}

\noindent {\it A comparison between CEM simulations of the early phases of the Milky Way and
observations of C, O and Fe gives indirect evidence that BH formation is rarely preceded by a (typical)
hypernova explosion where large amounts of $^{56}Ni$ are expelled, and that mass loss by stellar wind
during pre-WR phases and during the WR phase are heavy-metal-dependent}

\section{Stellar rotation and the effect on CEM's}\label{G}

\noindent The influence of rotation on massive single star evolution has been studied in detail by
Meynet and Maeder (2000) and by Heger et al. (2000). They showed that in
order to understand the evolutionary history and the chemistry of the outer layers of
individual stars, the effects of rotation are very important.

To investigate the chemical evolution of galaxies, an important
question is obviously in how far rotation affects average evolutionary properties.

From the studies listed above, we restrain the following evolutionary
effects.

\paragraph {The extend of the convective core}

The larger the rotation the larger the convective core during the core hydrogen
burning phase and this means that convective core overshooting mimics the effect of
rotation on stellar cores. The average equatorial velocity of OB-type stars $\sim$100-150
km/sec (Penny, 1996; Vanbeveren et al., 1998a, b). When this property is linked to the
calculations of Meynet and Maeder, we conclude that {\it the effect of rotation on the
extend of the convective core during the core hydrogen phase of massive single stars is on
the average similar to the effect of mild convective core overshooting as described in
Schaller et al. (1992)}.

Due to RLOF a binary component loses most of its hydrogen rich layers and the remnant 
corresponds more or less to the helium core left behind after core hydrogen burning.
Rotation makes larger helium cores so that also here convective core overshooting
mimics rotation.

Notice that the post-core hydrogen burning evolutionary phases are short, short enough to be able
to neglect the effects of rotation on the evolution of the extend of the convective core.

\paragraph {Mixing of the layers outside the convective core}

\noindent Rotation induces mild mixing in all mass layers outside the convective core
(rotational diffusion or meridional circulation) so that some matter of the convective core can be mixed
into the layers outside the convective core, even up to the outer observable envelope. Quantitatively
however, the overall chemical stellar yields, expelled by the star during its entire life, which serve
to enrich the ISM, hardly depend on this process. One exception: the formation of
primary nitrogen in zero metallicity stars (Meynet and Maeder, 2002).   

\paragraph {Rotation and stellar wind mass loss}

The evolution of massive stars is significantly affected by stellar wind mass loss, and not in the
least by the WR wind. Question: how does rotation affects the $\mdot$?  Using a theoretical hydrodynamic
atmosphere model Friend and Abbott (1986) proposed the following relation:

\begin{equation}
\mdot(\omega) = \mdot(\omega = 0)f(\omega)
\label{eq:rot}  		
\end{equation}

\noindent where $f(\omega)$ is an increasing function of the angular velocity $\omega$. Friend and 
Abbott themselves criticise their relation or using their own words {\it "...the
observations (their figure 4) show that there is no evidence for a dependence of the mass
loss rate on rotational velocity..."}. Even more: the relation of Friend and Abbott was seriously
questioned by Owocki et al. (1996), who performed hydrodynamic simulations of the winds of rotating hot
stars including the effect of nonradial radiation forces and gravity-darkening by using the von Zeipel
(1924) approximation. Something similar was done in detail by Petrenz and Puls (2000) who
demonstrated that with a correct NLTE treatment of the outflowing envelope of a rotating star, the
$\mdot$ differs by at most 10-20 $\%$ from the $\mdot$ of the same star but non-rotating, even
if the star is rotating close to break-up.

Yet, an expression like Equation~\ref{eq:rot} is used by some scientists and included in their stellar
evolutionary code (Maeder and Meynet, 2000; Heger et al., 2000). As a matter of fact, since rotation
hardly affects the stellar interior structure during the post-core-hydrogen burning phases, the main
effect of rotation on stellar evolution during these post-core-hydrogen burning phases is primarily
due to the adopted Equation~\ref{eq:rot} to describe the $\mdot$ and the effect of
rotation on the latter. Ignoring for the moment the possible physical incorrectnes, 
it is important to realise that Equation~\ref{eq:rot} is a relation between the $\mdot$ of a
non-rotating star and the $\mdot$ of the same star which is rotating. So, if
Equation~\ref{eq:rot} is used in an evolutionary code, what kind of an expression must be used to
describe the $\mdot$ of a non-rotating star? Maeder and Meynet and Heger et al. replace the latter
by existing empirical $\mdot$ formalisms (the formalism of Nieuwenhuijzen and de Jager, 1990
for the pre-WR evolutionary phases, the one of Langer, 1989 or of NL during the WR phase). Since these
empirical formalisms include data from stars that are rotating (most of them), the way Maeder and
Meynet and Heger et al. study the effect of rotation on stellar wind mass loss may be wrong.

Obviously, a better formalism would be a formalism where the mass loss rate data are derived by an
atmosphere model that includes the effects of rotation. This was done for a few stars by Petrenz and
Puls (1996). An important conclusion of the latter study is that {\it all the mass loss rates that are
determined from the H$\alpha$ line profiles, are overestimated typically by 20-30
$\%$ if the effects of rotation are not included}. This means that the empirical mass loss rate
formalisms that were proposed in the past (and used in some stellar evolutionary codes) may
overestimate the effects of stellar wind mass loss.

\paragraph{Rotation and CEMs}

\noindent Accounting for the discussion on the effects of rotation on stellar evolution in the previous
paragraphs, we have made several CEM simulations with stellar chemical yields which include AVERAGE
effects of rotation that are discussed above and we have to conclude that {\it the effect of rotation on
general results of chemical evolutionary models of galaxies is similar to the effect of moderate
convective core overshooting and is much smaller than the overall effects of binaries, with one possible
exception: the formation of primary nitrogen during the very early evolution of galaxies}. Since most
of the present CEMs use stellar evolutionary calculations with moderate amount of convective core
overshooting, most of the CEMs already accounted for stellar rotation {\it avant la lettre}. Of course,
there may still be evolutionary effects which have not been recognised yet.

\end{document}